\documentclass[12pt, preprint]{aastex}

\usepackage{graphicx}
\usepackage{amssymb}
\usepackage{layout}
\usepackage{url}
\usepackage{subfigure}
\usepackage{color}
\usepackage{enumerate}
\slugcomment{Submitted to {\it The Astrophysical Journal}}

\begin{document}


\title{Polarization Signatures of Kink Instabilities in the Blazar Emission Region from Relativistic Magnetohydrodynamic Simulations}


\author{Haocheng Zhang\altaffilmark{1,2}, Hui Li\altaffilmark{2}, Fan Guo\altaffilmark{2}, and Greg Taylor\altaffilmark{1}}

\altaffiltext{1}{Department of Physics and Astronomy, University of New Mexico, Albuquerque, NM 87131, USA}

\altaffiltext{2}{Theoretical Division, Los Alamos National Laboratory, Los Alamos, NM 87545, USA}

\begin{abstract}
Kink instabilities are likely to occur in the current-carrying magnetized plasma jets. Recent observations of the blazar radiation and polarization signatures suggest that the blazar emission region may be considerably magnetized. While the kink instability has been studied with first-principle magnetohydrodynamic (MHD) simulations, the corresponding time-dependent radiation and polarization signatures have not been investigated. In this paper, we perform comprehensive polarization-dependent radiation modeling of the kink instability in the blazar emission region based on relativistic MHD (RMHD) simulations. We find that the kink instability may give rise to strong flares with polarization angle (PA) swings or weak flares with polarization fluctuations, depending on the initial magnetic topology and magnetization. These findings are consistent with observations. Compared with the shock model, the kink model generates polarization signatures that are in better agreement with the general polarization observations. Therefore, we suggest that kink instabilities may widely exist in the jet environment, and provide an efficient way to convert the magnetic energy and produce multiwavelength flares and polarization variations.
\end{abstract}

\section{Introduction}

Relativistic jets are common in many high-energy astrophysical systems, such as active galactic nuclei (AGNs), gamma-ray bursts (GRBs), X-ray binaries, etc.. Among these systems, blazars, which are a type of AGNs whose jets direct very close to our line of sight (LOS), provide a unique opportunity for jet study, as they shed light on both the global and inner-jet properties.

Blazars emit nonthermal-dominated radiation covering the entire electromagnetic spectrum from radio up to TeV $\gamma$-rays. Blazar spectral energy distributions (SEDs) have two broad, non-thermal components. The low-energy component, from radio to optical-UV, is believed to be synchrotron emission by ultrarelativistic electrons in a partially ordered magnetic field. This is evident by the observed high polarization degree \citep[PD; see for example,][]{Scarpa97}. The high-energy component, from X-rays to $\gamma$-rays, is due to either the inverse Compton scattering by the nonthermal electrons that make the low-energy component \citep[leptonic model; e.g.,][]{Marscher85,Dermer92,Maraschi92,Sikora94}, or the synchrotron by ultrarelativistic protons and cascading secondary particles \citep[hadronic model; e.g.,][]{Mannheim92,Mucke01}. So far the two models cannot be distinguished, but several diagnostics such as the neutrino flux, multiwavelength variability, and high-energy polarization have been put forward \citep[e.g.,][]{Halzen97,ZHC13,Diltz15,Petropoulou15,ZHC16b}. Both components show strong variability, with time scales ranging from minute-scale flares to month- or even year-scale active phases \citep[e.g.,][]{Ostorero04,Aharonian07}. For individual flares, most of them last between hours to a few tens of days. Based on the causality relation, the size of the emission region should be $R\lesssim \delta c t_v$, where $\delta$ is the Doppler factor and $t_v$ is the observed variability time scale, $R\lesssim 1~\rm{pc}$ is relatively small compared to the size of the jet. Therefore, it is often believed that the blazar emission comes from an unresolved region along the jet. Moreover, both the radio and optical polarization signatures appear variable as well \citep[e.g.,][]{Marscher08,Darcangelo09,Covino15,Blinov16}. In general, the optical PD fluctuates around $10-20\%$; in very rare cases it can go up to $40-50\%$. The optical PA usually fluctuates around a mean value, but sometimes it can make $\sim 180^{\circ}$ swings \citep[e.g.,][]{Abdo10,Blinov15,Kiehlmann16}. While some of these PA swings appear to be random walks, which are consistent with a turbulent model \citep{Marscher14}, others are deterministic (not random), which can be explained by the magnetic field evolution or the LOS variations \citep{Marscher08,Abdo10,ZHC15}. In addition, some of these deterministic PA swings are correlated with multiwavelength flares \citep{Marscher08,Abdo10,Blinov15}.

In order to understand the jet emission, the knowledge of jet energy composition and the particle acceleration mechanism is essential. In the case of blazars, shock models have been widely used to explain the flaring activities \citep[e.g.,][]{Marscher85,Spada01,Joshi07,Graff08}. Shock models generally assume that the emission region has a significant amount of kinetic energy, which can be converted to nonthermal particle energy through the shock acceleration. In a low magnetization environment, shock compression ratio is high, leading to a power-law nonthermal particle spectrum of index $\gtrsim 2$ \citep{Achterberg01,Spitkovsky08,Summerlin12}. This is consistent with common blazar SEDs \citep[e.g.,][]{Boettcher13,Cerruti15}. Additionally, the observed standing and moving radio knots can also be explained by shock models \citep[e.g.,][]{Biretta95,Lobanov99,Venturi99}.

However, polarization signatures pose some questions on the shock model. Strong shocks lead to strong compression of the magnetic field in the emission region, especially in a low magnetization environment, which is typical for efficient shock acceleration. Through detailed polarization-dependent radiation simulations based on RMHD simulations, \cite{ZHC16a} have shown that in the case of a laminar shock layer, the PD goes beyond $40-50\%$ during flares if the flare level is more than a doubling. Therefore, in order to keep the PD within the observed range of $10-20\%$, either the shock acceleration cannot be the main source for nonthermal particles, or the shock should appear in multiple directions such as in a turbulent model. The latter, however, cannot interpret the deterministic PA swings. In addition, some recent observations indicate very hard SEDs during flares \citep[e.g.][]{Hayashida15}, which can hardly be obtained by the standard shock model \citep[e.g.,][]{Kirk00,Achterberg01}.

If the jet emission region has a considerable amount of magnetic energy, kink instability may lead to substantial energy release and particle acceleration through the dynamical evolution and its associated magnetic reconnection. Several papers have demonstrated how the local magnetic field evolves during kinks in the jet environment, and that the global jet can still keep collimated \citep[e.g.,][]{Mizuno09,Guan14}. Additionally, other magnetic energy conversion processes can happen along with kinks, such as magnetic reconnection. Particle-in-cell (PIC) simulations have shown that reconnection can make power-law nonthermal particle spectra \citep[e.g.,][]{Sironi14,Guo14,Guo16,Werner16}. In addition, both RMHD and PIC simulations have illustrated possible radiation and polarization signatures resulting from magnetic reconnection \citep[e.g.,][]{Deng16,Yuan16}. Nevertheless, the radiation and polarization signatures that are intrinsically from kink instabilities are so far not well studied.

In this paper, we present comprehensive radiation and polarization simulations of the kink instability in the blazar emission region, based on RMHD simulations. Our approach is self-consistent in the magnetic field evolution, and provides detailed analysis on the evolution of the kink instability and how it is related to the observational features. This enables us to constrain the physical parameters in the blazar emission region by comparing our results with the general observational phenomena. The paper is organized as follows: we will describe our model and simulations in Section \ref{model}, illustrate the kink evolution and the resulting polarized emission, and implications for observations in Sections \ref{result} and \ref{implication}, and discuss our results in Section \ref{discussion}.

\section{Model Description \label{model}}

Our goal is to study the radiation and polarization signatures that intrinsically come from the kink instability in the blazar emission region. In addition, we want to illustrate how the observables are affected by the magnetization and the initial magnetic topology in the emission region. In order to facilitate direct comparisons, we try to employ the simplest physical assumptions. In the following, we will describe our physical assumptions and the simulation setup.

\subsection{Physical Assumptions}

We perform our simulations in the blazar emission region. Generally speaking, since the blazar emission region is an unresolved region along the blazar jet, the magnetic field and the plasma evolution in the emission region should be influenced by the large-scale jet evolution. However, the observed fast variability and high luminosity of the emission region suggest that the emission region is an extraordinary, localized region that evolves very fast compared to the global jet evolution. Therefore, we make the assumption that within the time scale of interest in this work, the evolution of the emission region can be treated to be detached from the large-scale jet evolution. For the magnetic topology in the emission region, several papers have shown that the observed polarization signatures are consistent with a helical magnetic field \citep{Lyutikov05,Pushkarev05,ZHC15}, with a possible turbulent component. Here we focus on features from kink instabilities, hence we choose a laminar initial setup, although some turbulence is expected to develop in the nonlinear stage of the kink instability. Usually we are observing blazars very close to the jet direction ($\theta_{\rm obs}^{\star}\sim1/\Gamma$) in the observer's frame, equivalent to $\theta_{\rm obs} \sim 90^{\circ}$ in the comoving frame of the jet. As is suggested by the bending jet scenario, a change in the LOS ($\theta_{\rm obs}^{\star}$) can considerably alter the observed radiation and polarization signatures \citep[e.g.,][]{Abdo10,Kiehlmann16}, this effect is however due to the geometry rather than the kink instability. Thus we fix our LOS at $\theta_{\rm obs}=90^{\circ}$ in the comoving frame, so that the Doppler factor $\delta \equiv \left( \Gamma \, [1 - \beta_{\Gamma} \, \cos\theta_{\rm obs}^{\star}] \right)^{-1} \sim \Gamma$.

Our model assumes that the emission region is a cylindrical region traveling along the jet with bulk Lorentz factor of $\Gamma=20$ in the observer's frame. In the comoving frame of the emission region, it is pervaded by a helical magnetic field. The kink instability sets off if the Kruskal-Shafranov criterion, also called the safety factor, $q=\frac{2\pi r}{L}\frac{B_z}{B_{\phi}}$, is smaller than one. We envision a situation that the magnetic fields within the emission region are accumulated with the large-scale jet evolution. Here we start the simulation with a given magnetization and a specific magnetic topology, which is already unstable to the kink. The starting helical magnetic field is in a magnetic force balance. In this way, the plasma density and the pressure can be initialized to be uniform in the simulation box. We assume that in the beginning the plasma is cold, but we can adjust plasma density to vary the initial magnetization in the emission region. We put a small perturbation in the velocity field to trigger the kink instability.

In view of the complicated nonthermal particle acceleration and evolution processes, which cannot be self-consistently handled by MHD simulations and are beyond the scope of this paper, we simplify the nonthermal particle population as two components. One is a constant component of background particles that are uniformly distributed in the entire emission region. These particles may originate from the stochastic acceleration by the microscopic turbulence that is not explicitly considered in the RMHD simulations. These particles are necessary to make the quiescent state of the blazar emission. The other is the injected particles (flare particles) due to the conversion of the magnetic energy induced by the kink instability. These particles may be accelerated through magnetic reconnection that arises due to the bending jet during the kink evolution \citep{Singh16}. We assume that the energetic particles are isotropic, and the injection rate is a constant portion of the magnetic energy conversion rate induced by the kink instability, using the local $\vec{j}\cdot \vec{E}$ (where $\vec{E} \sim -\vec{v} \times \vec{B}/c$) to normalize the nonthermal particle injection rate locally. Additionally, details of the stochastic acceleration and the magnetic energy conversion through the kink instability are beyond the scope of this paper. Here we make the simple assumption that the two components share the same spectral shape. Finally, for bright blazars the synchrotron and Compton scattering cooling in the blazar emission region is strong, so that we can simplify the radiative cooling by replacing the nonthermal particles each time step, assuming that most particles have been sufficiently cooled.

We summarize our physical assumptions in the following:
\begin{enumerate}
\item The emission region is a localized cylindrical region whose evolution is not linked to the large-scale jet;
\item We are observing the emission region along a fixed LOS, at $\theta_{\rm obs}=90^{\circ}$ in the comoving frame of the emission region.
\item We do not simulate the energy accumulation process of the system; instead, we start the simulation with a given magnetization and a magnetic topology that are unstable to the kink instability.
\item The initial magnetic field is a helical force-free field, and the plasma density and the thermal pressure are uniform.
\item The plasma is cold in the beginning.
\item A velocity perturbation is set to stimulate the kink instability.
\item All flow conditions and the magnetic field start as laminar.
\item Every time step we put in a constant background nonthermal particle component and an injected particle component that is normalized by the magnetic energy converted through the kink instability; they are replaced each time step to mimic the cooling process.
\end{enumerate}

\subsection{Simulation Setup}

\begin{figure}[ht]
\centering
\includegraphics[width=\linewidth]{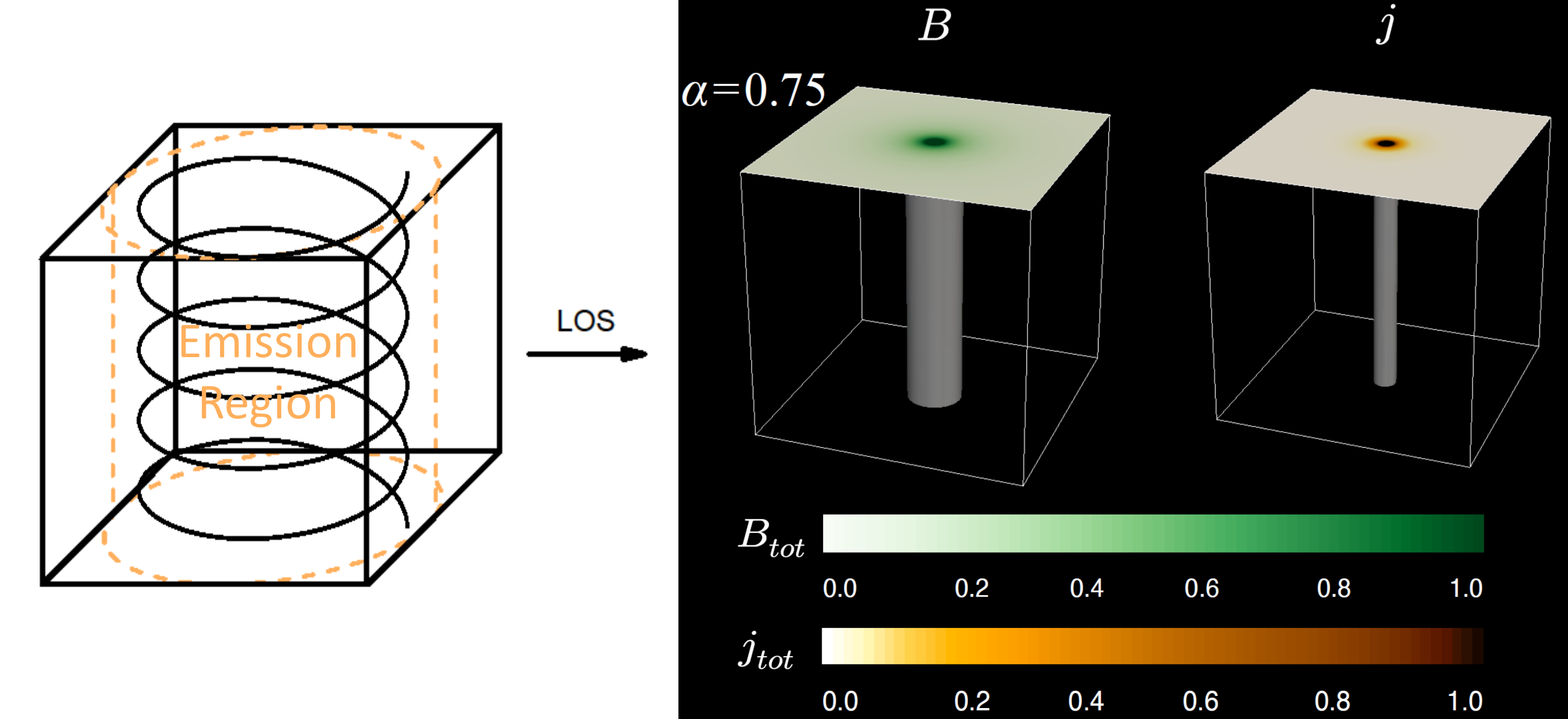}
\caption{Initial setup. Left: a sketch of the simulation box. The simulation box is pervaded by a helical magnetic field, whose strength decreases along the radial direction. Most of the emission comes from the orange emission region. LOS is to the right in the comoving frame of the emission region. Refer to our model description for details. Right: 3D isosurfaces of the magnetic field strength and the magnitude of the current density, both with transverse slice at the top of the simulation box. \label{setup}}
\end{figure}

Our model is realized by coupling the 3D multi-zone RMHD code LA-COMPASS developed by \cite{Li03} and the 3D multi-zone polarization-dependent ray-tracing code 3DPol by \cite{ZHC14}. Fig. \ref{setup} shows our initial setup of the RMHD simulation, and Table \ref{table} lists the conversion between the RMHD code units and the physical units. LA-COMPASS is performed in the comoving frame of the emission region, using Cartesian coordinates. The simulation box takes outflow boundaries, except along the z-axis, where we use the periodic boundary. This is to prevent plasma moving in and out of the z-boundary when the kink grows, and to ensure the closure of the magnetic field lines. This arrangement is also in accordance with the fact that the emission region is embedded in the global jet. The simulation box is a cube, with x, y, and z range from $-L$ to $L$, where $L=16$. We take the force-free helical magnetic field setup from \cite{Mizuno09}, in the form of 
\begin{equation}
\begin{array}{c}
B_z=\frac{B_0}{(1+(r/r_0)^2)^{\alpha}} \\
B_{\phi}=\frac{B_0}{(r/r_0)(1+(r/r_0)^2)^{\alpha}}\sqrt{\frac{(1+(r/r_0)^2)^{2\alpha}-1-2\alpha(r/r_0)^2}{2\alpha-1}}
\end{array}
\end{equation}
where $r$ is the radial distance from the central axis of the cylindrical emission region, $B_0$ parameterizes the magnetic field strength, $r_0$ characterizes the radius of the region with significant magnetic energy, and $\alpha>0.5$ controls the ratio between the toroidal and poloidal magnetic components. Though the total toroidal component dominates for all $\alpha$ values, smaller $\alpha$ strengthens the initial poloidal component. $B_r$ is set to be zero initially. The magnetization factor is defined as
\begin{equation}
\sigma = \frac{E_{em}}{h}
\end{equation}
where $E_{em}=\frac{B^2+E^2}{8\pi}$ is the electromagnetic energy density, $h=\rho c^2+\frac{\hat{\gamma} p}{\hat{\gamma}-1}$ is the specific enthalpy, $\rho$ is the plasma density, $\hat{\gamma}$ is the adiabatic index and $p$ is the thermal pressure. The plasma is assumed to be cold, so that $p$ is set to be $0.25$ (see Table \ref{table} for physical units). $\sigma$ is adjusted through the plasma density $\rho$. Both the magnetic field strength and $\sigma$ have dependence on $r$. Thus we use the maximal magnetization factor ($\sigma_m$) at the central axis of the emission region and $B_0$ to normalize to physical values. The initial velocity perturbation is given by
\begin{equation}
v_r=0.01\times e^{-\frac{r}{r_0}}\cos(\theta)\sin(\frac{2\pi nz}{L})
\end{equation}
where $\theta$ is the azimuthal angle. We choose $n=4$ rather than $n=1$ as in \cite{Mizuno09}, because in the case of $n=1$ the kink develops too fast to study the details in the radiation features. The background nonthermal particle density ($n_{bkg}$) and the injected nonthermal particle density ($n_{inj}$) take the spectral shape of
\begin{equation}
\begin{array}{c}
n_{bkg}=n_0\times \gamma^{-2} \\
n_{inj}=Q_0\times \gamma^{-2}
\end{array}
\end{equation}
where $\gamma$ is the Lorentz factor of the nonthermal electrons, ranging from 1 to $10^5$, and $Q_0$ is normalized by half of the magnetic energy conversion rate $\vec{j}\cdot \vec{E}$ during the kink evolution, where $\vec{j}=\frac{\nabla \times \vec{B}}{c}$, and $\vec{E}=-\frac{\vec{v}\times\vec{B}}{c}$. The multi-zone time-dependent magnetic field and the nonthermal particle information will be fed into the 3DPol code.

The 3DPol code calculates the time-dependent radiation and polarization signatures of synchrotron emission using ray-tracing method through the addition of Stokes parameters. This method naturally includes all light travel time effects (LTTEs). Details of the code capability can be found in \cite{ZHC14,ZHC15}. The viewing angle in the comoving frame is fixed along the y-axis. Here we only consider the synchrotron emission in the optical band, so that any synchrotron-self absorption or Faraday rotation effects are negligible. The 3DPol code is performed in the comoving frame of the emission region as well, but it will Lorentz transform the final time-dependent radiation and polarization signatures into the observer's frame at the end of the simulation. We define the PA in the observer's frame in the following way. When the electric vector is parallel to the emission region propagation (toroidal component dominating), ${\rm PA}=0$. However, during the kink instability, the meaning of poloidal and toroidal components does not directly reflect the magnetic field direction. Therefore, we use vertical component for the magnetic field lying parallel to the jet propagation direction, and use planar component for the perpendicular direction. Since the PA has $180^{\circ}$ ambiguity, the planar (initially toroidal) domination happens at ${\rm PA}=2N\times90^{\circ}$, and the vertical (initially poloidal) domination happens at ${\rm PA}=(2N+1)\times90^{\circ}$, where $N$ is an integer.

\begin{table}[ht]
\scriptsize
\parbox{\linewidth}{
\centering
\begin{tabular}{|l|c|c|c|}\hline
Parameters     & Length               & Time               & Velocity                  \\ \hline
Relation       & $L_0$                & $L_0/c$            & $c$                      \\ \hline
Code Unit      & $1$                  & $1$                & $1$                      \\ \hline
Physical Value & $3\times10^{16}~{\rm cm}$ & $1\times10^6~{\rm s}$   & $3\times10^{10}~{\rm cm~s^{-1}}$    \\ \hline
Parameters     & Magnetic Field       & Thermal Pressure   & Plasma Density            \\ \hline
Relation       & $B_0$                & $B_0^2/(4\pi)$     & $B_0^2/(4\pi c^2)$       \\ \hline
Code Unit      & $1$                  & $1$                & $1$                      \\ \hline
Physical Value & $0.3~{\rm G}$ & $7.0\times10^{-3}~{\rm erg~cm^{-3}}$ & $8.0\times10^{-24}~{\rm g~cm^{-3}}$ \\ \hline
\multicolumn{4}{c}{}\\
\end{tabular}}
\caption{Conversion between the RMHD code units and the physical units. All parameters are in the comoving frame. Considering a bulk Lorentz factor $\Gamma=20$, one code time unit is about $0.5~{\rm d}$. \label{table}}
\end{table}

\section{Results \label{result}}

In this section, we investigate the magnetic energy conversion during the kink instability in the blazar emission region, and how it affects the time-dependent radiation and polarization signatures. \cite{Mizuno09} have found that the kink evolution is governed by the plasma density distribution and the magnetic topology. The former is essentially the effect of the magnetization. Here we start with uniform plasma density for all cases, but vary it to make different initial magnetization. During the development of the kink, the magnetic energy keeps converting to the kinetic and the thermal energies, so that the magnetization factor $\sigma$ varies in time. Furthermore, the magnetic topology changes in time as well. Therefore, we choose the initial $\sigma_m$ and $\alpha$ to describe the system. We present three cases, namely, $\sigma_m=2$ and $\alpha=0.75$ (Case 1), $\sigma_m=2$ and $\alpha=1$ (Case 2), and $\sigma_m=0.2$ and $\alpha=0.75$ (Case 3). The RMHD simulations are shown in Figs. \ref{case1}, \ref{case2}, and \ref{case3}, respectively, along with the magnetic energy evolution and the injection level in Fig. \ref{kink1}, and finally the time-dependent radiation and polarization signatures in Fig. \ref{kink2}. For all these cases, $n_{bkg}$ and $B_0$ are fixed in order to keep a similar quiescent state and facilitate direct comparison.

The initial velocity perturbation can generate multiple kink modes. One of these modes grows the fastest, and generally dominates the kink evolution. The kink evolution can be characterized as two epochs: the first epoch is called the linear growth phase, where the fastest kink mode grows exponentially in time; the second epoch is called the nonlinear growth phase, where the fastest kink growth becomes saturated, and the slower kink modes may continue to grow and convert the magnetic energy (Fig. \ref{kink1}). The whole evolution is similar to \cite{Mizuno09}. Specifically, during the linear growth phase, even though the kink instability grows exponentially, its growth has not become significant, so that the magnetic energy conversion is relatively low. We call this period Phase A. Phase B refers to the rest of the linear growth phase, where the magnetic energy conversion is strong. At the beginning of the nonlinear growth phase, since our simulation lasts longer than those in \cite{Mizuno09}, we observe multiple slower kink modes grow up, which continue to convert the magnetic energy (Fig. \ref{kink1} upper panel). We call this period Phase C. The rest of the nonlinear growth phase is called Phase D, where the kink evolution becomes very slow.

As we use the same nonthermal particle spectrum for both $n_{bkg}$ and $n_{inj}$ without any explicit spectral evolution, we only study the light curves and polarization signatures in the optical band. We use the relative flux level here, where the quiescent state flux $\sim 1$ is approximately $10^{45}~ {\rm erg\,s^{-1}}$, corresponding to a bright blazar such as 3C~279. As we can see in the following that the presence of a quiescent state influences the radiation and polarization signatures from the kink instability, the exact value however does not alter the general trend.

\subsection{Kink and Polarization-dependent Radiation}

\begin{figure}[ht]
\centering
\includegraphics[width=\linewidth]{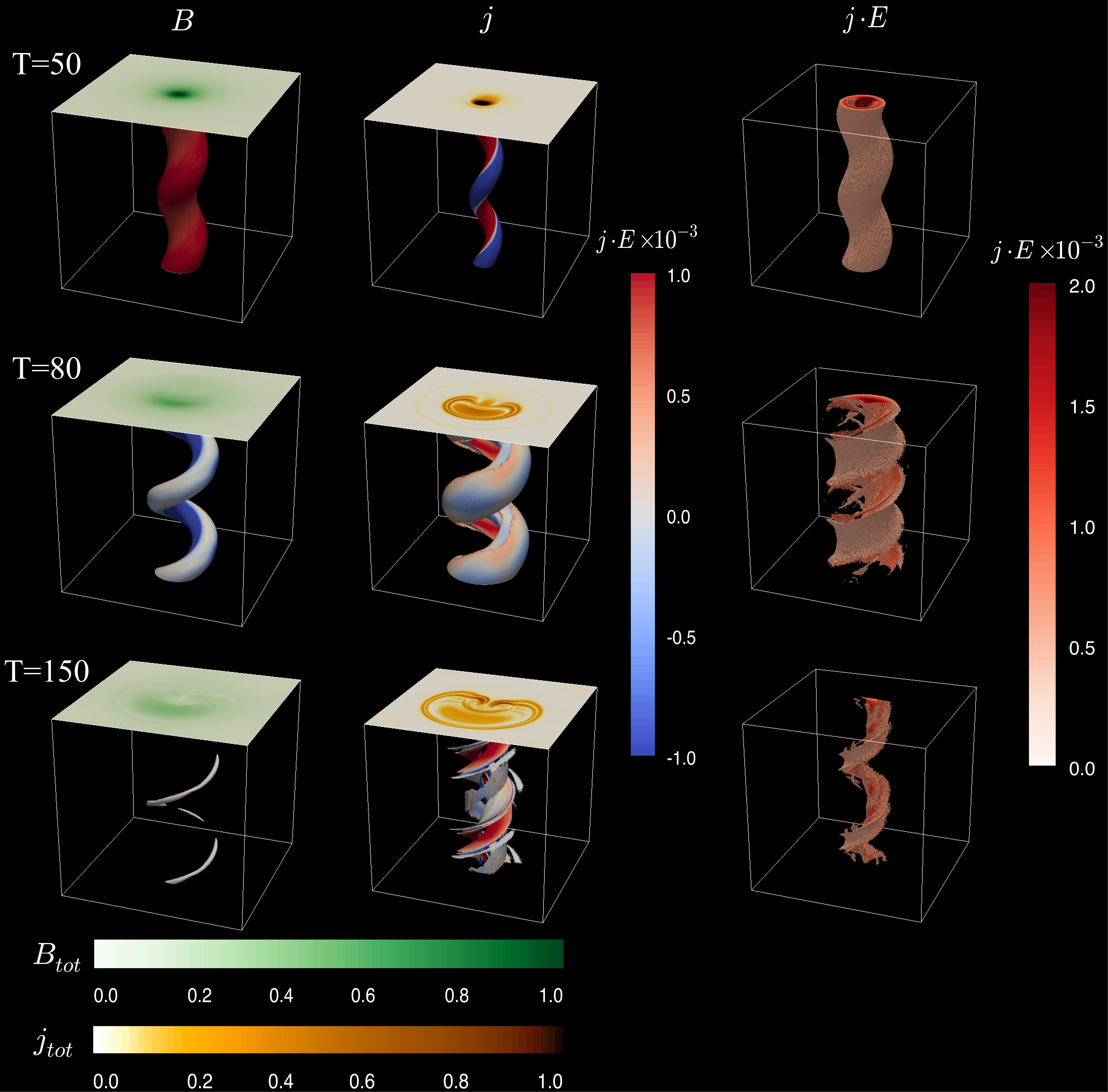}
\caption{The kink evolution for Case 1. Left column is the 3D isosurface of the magnetic field strength at $B=0.2$, with transverse slice at the top of the simulation box. Middle column is those for the magnitude of the current density, with the isosurface chosen at $|\vec{j}|=0.2$. The color on both isosurfaces present the distribution of $\vec{j}\cdot \vec{E}$. Right column plots all zones with a positive $\vec{j}\cdot \vec{E}$, and the color indicates the strength of $\vec{j}\cdot \vec{E}$. Panels are selected at code units $T=50$ (upper row), $T=80$ (middle row), and $T=150$ (lower row). \label{case1}}
\end{figure}

We first take Case 1 as an example to illustrate effects of the kink evolution on the radiation and polarization signatures. In this case, within the region where the magnetic field is large, the average magnetization is $\sigma\sim 1$, so that the magnetic field is likely to actively participate in the system evolution and provide energy. This is clearly shown in the simulation. During Phase A ($0<T<50$ see Table \ref{table} for physical units), the magnetic energy conversion is not strong. We notice that at $T\sim 20$, the positive $\vec{j}\cdot \vec{E}$ starts to increase, but it is balanced by a negative $\vec{j}\cdot \vec{E}$, corresponding to the pressure outside the magnetic region that prevents it from expanding. This balance is broken when the kink evolves to Phase B ($50<T<80$) at $T\sim 50$. We can see in Fig. \ref{case1} (upper panel) that regions with positive $\vec{j}\cdot \vec{E}$ becomes dominating. We notice that these regions concentrates around the central axis of the system, where the vertical magnetic component is strong. Additionally, the positive $\vec{j}\cdot \vec{E}$ peaks at the beginning of Phase B (Fig. \ref{kink1} lower panel). During this phase, the kink instability quickly converts the magnetic energy and bends the magnetic field and current into helices (Fig. \ref{case1} upper and middle panels), then it evolves into the nonlinear phase at $T\sim 80$. However, even during this period of fast kink evolution, the kink instability does not generate relativistic bulk flows. This is probably because the magnetization is not adequately high in our simulations. The positive $\vec{j}\cdot \vec{E}$ is still considerable at Phase C ($80<T<150$), but it moves away from the central axis, where the planar magnetic component becomes stronger (Fig. \ref{case1} right column of the middle and lower panels). By comparing Fig. \ref{case1} upper and middle panels with the lower panel, we notice that the first two panels exhibit a smooth helical structure made by the fastest growing kink mode, but the lower panel shows that the structure is more complex. This suggests that the slower growing kink modes give rise to some turbulence, which roughens the smooth helical structure. Nevertheless, the turbulence is relatively weak, so that the general helical structure is still kept. These slower kink modes also explains the few major magnetic energy drops during Phase C (Fig. \ref{kink1} upper panel). Nevertheless, their contribution to the positive $\vec{j}\cdot \vec{E}$ is small compared to the fastest mode. $T\sim 150$ marks the beginning of Phase D ($150<T<250$). While it is clear that the system is still evolving, the general magnetic topology variation and magnetic energy conversion become trivial. Thus the following evolution is of little interest.

The radiation and polarization signatures follow the above evolution. Since the magnetic field is dominated by the toroidal (planar) component in the beginning, the PD is at a relatively high level, $\sim 35\%$, and PA rests at $180^{\circ}$. When the kink starts to grow, the flux level increases. As the injection is proportional to the positive $\vec{j}\cdot \vec{E}$, which initially concentrates around the central axis where the vertical magnetic component is stronger, the PD drops towards zero and the PA rotates towards $90^{\circ}$, corresponding to the transition to the vertical component domination. The flux peaks at $t\sim 30~{\rm d}$, corresponding to the injection peak at Phase B, $T\sim 50$ ($t\sim 25~{\rm d}$). This is also evident by comparing the flare level and injection level, both of which are about two times higher than the background (Figs. \ref{kink1} and \ref{kink2}). The $\sim 5~{\rm d}$ delay of the flare peak compared to the injection peak is due to the LTTEs. At this epoch, the emission is dominated by the strong injection at central regions dominated by the vertical magnetic component. Thus the PD rises to a small peak at the flare peak, and PA stays around $90^{\circ}$. At the end of the flare, the magnetic field is sufficiently altered by the kink instability. Although the planar contribution is still stronger than the vertical, the planar dominance is less strong. Therefore, we observe that the PD rests at $\lesssim 20 \%$. In addition, the flux level is lower than the initial state, as the magnetic energy now is lower. We can see that the final flux level is about $80 \%$ of the initial state, in agreement with the final magnetic energy percentage.

\subsection{Effects of the Initial Topology}

We now examine the influence of the initial field topology. Case 2 has a stronger toroidal component than Case 1 in the beginning. Therefore, the initial PD is higher than Case 1, at $\sim 45\%$ (Fig. \ref{kink2}). In addition, the initial magnetic field distribution is different from Case 1: the magnetic field concentrates more towards the central axis of the emission region (Fig. \ref{setup}), and the total magnetic energy is slightly lower. Hence we can see that during the initial quiescent state, Case 2 has a little bit lower flux level than Case 1 (Fig. \ref{kink2}).

\begin{figure}[ht]
\centering
\includegraphics[width=\linewidth]{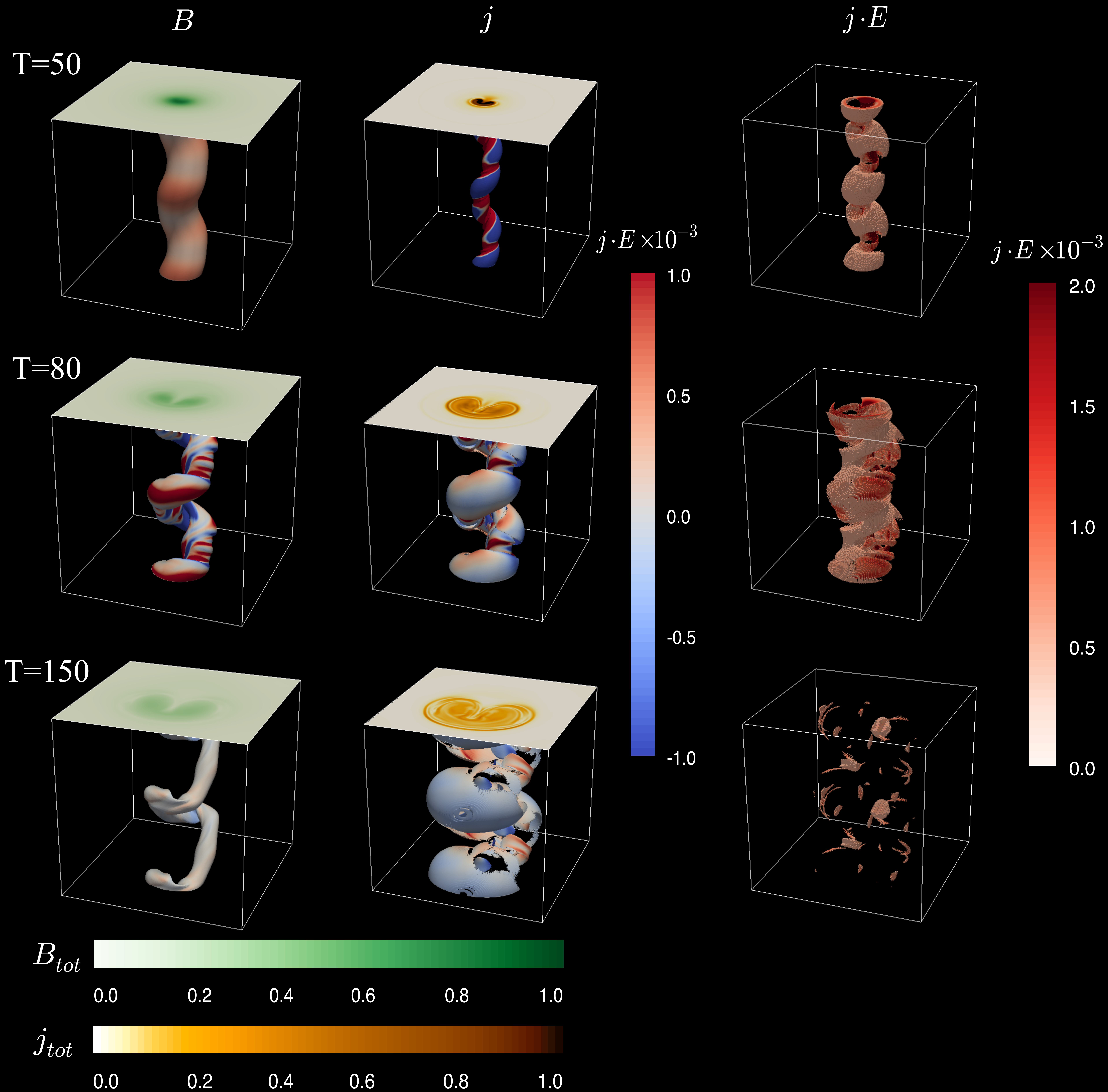}
\caption{The kink evolution for Case 2. The magnetic field and the current isosurfaces are chosen at $B=0.15$ and $|\vec{j}|=0.15$, respectively. Otherwise all panels, plots, and color maps are the same as those in Fig. \ref{case1}. \label{case2}}
\end{figure}

\begin{figure}[ht]
\centering
\includegraphics[width=0.7\linewidth]{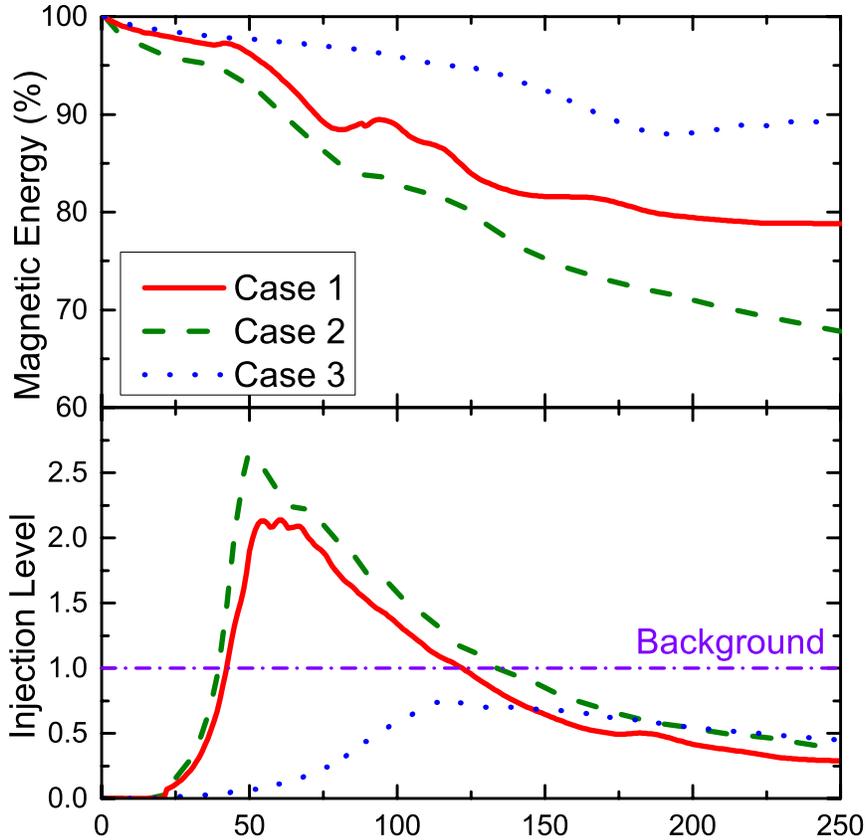}
\caption{Time-dependent magnetic energy and the injection level for all cases in the code time units. Upper panel shows the percentage of the initial magnetic energy in time. Lower panel shows the injected nonthermal particle density ($n_{inj}$) level , which is half of the average positive $\vec{j}\cdot \vec{E}$ normalized by $n_{bkg}$ in time. Red solid lines are for Case 1, green dashed lines are for Case 2, and blue dotted lines are for Case 3. The purple dot-dashed line indicates $n_{bkg}$. The total nonthermal particle density is $n=n_{inj}+n_{bkg}$ for each zone. \label{kink1}}
\end{figure}

The stronger toroidal component makes the emission region more unstable to the kink. Fig. \ref{kink1} shows that the magnetic energy conversion rate is higher than Case 1, also more magnetic energy is converted through the kink. This leads to stronger nonthermal particle injection, so that we can observe the flare level is higher than Case 1 (Fig. \ref{kink2}). Although the flares in the two cases peak at the same time, we can observe in Case 2 that the positive $\vec{j}\cdot \vec{E}$ distributes farther away from the central region than that in Case 1 (Fig. \ref{case2} right column), thus the vertical component dominance is less significant in Case 2. Hence, we can see that at the flare peak, the PD rises to a lower value than in Case 1, and the PA swing appears smoother. At the end of the flare, the flux level is lower than that in Case 1, as more magnetic energy is released during the kink.

Nevertheless, Case 2 is unlikely to happen in reality. The highest observed blazar PD is around $40-50 \%$ \citep{Scarpa97,Lister05}, so that the initial PD in Case 2 is too high, although a strong turbulent component may lower the initial PD. However, since Case 2 is very unstable to kink, a strong turbulent component may have already triggered the instability before the toroidal component grows too strong. Consequently, based on the PD that is frequently observed, we suggest that the initial magnetic topology for the kink instability should generally have comparable toroidal and poloidal components in the comoving frame. In this situation, the emission region mostly has a safety factor around 1, which permits some time to accumulate the initial magnetic energy before triggering the kink instability.

\subsection{Effects of the Initial Magnetization}

\begin{figure}[ht]
\centering
\includegraphics[width=\linewidth]{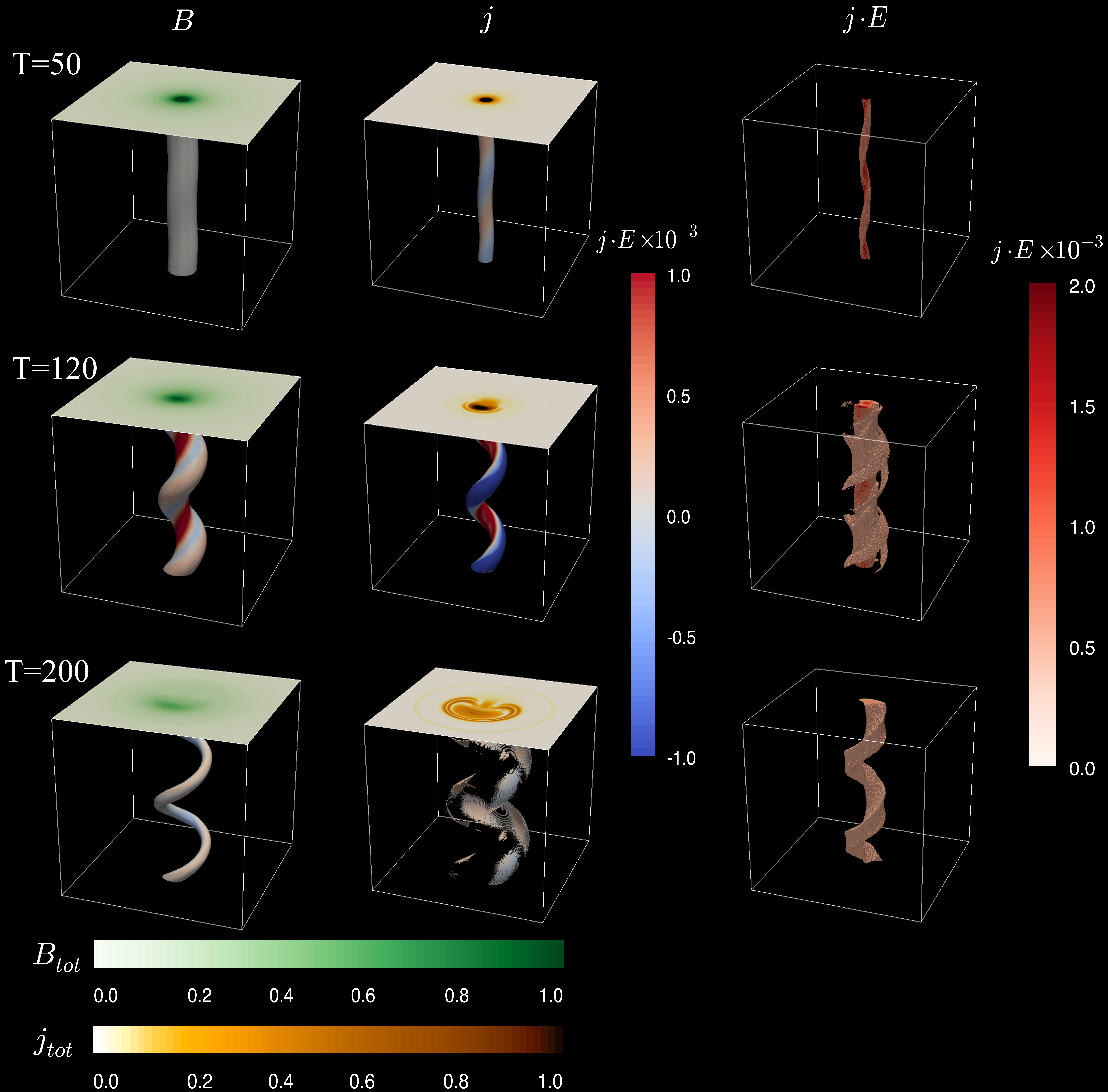}
\caption{The kink evolution for Case 3. The magnetic field and the current isosurfaces are chosen at $B=0.25$ and $|\vec{j}|=0.25$, respectively. Otherwise all panels, plots, and color maps are the same as those in Fig. \ref{case1}.\label{case3}}
\end{figure}

\begin{figure}[ht]
\centering
\includegraphics[width=0.7\linewidth]{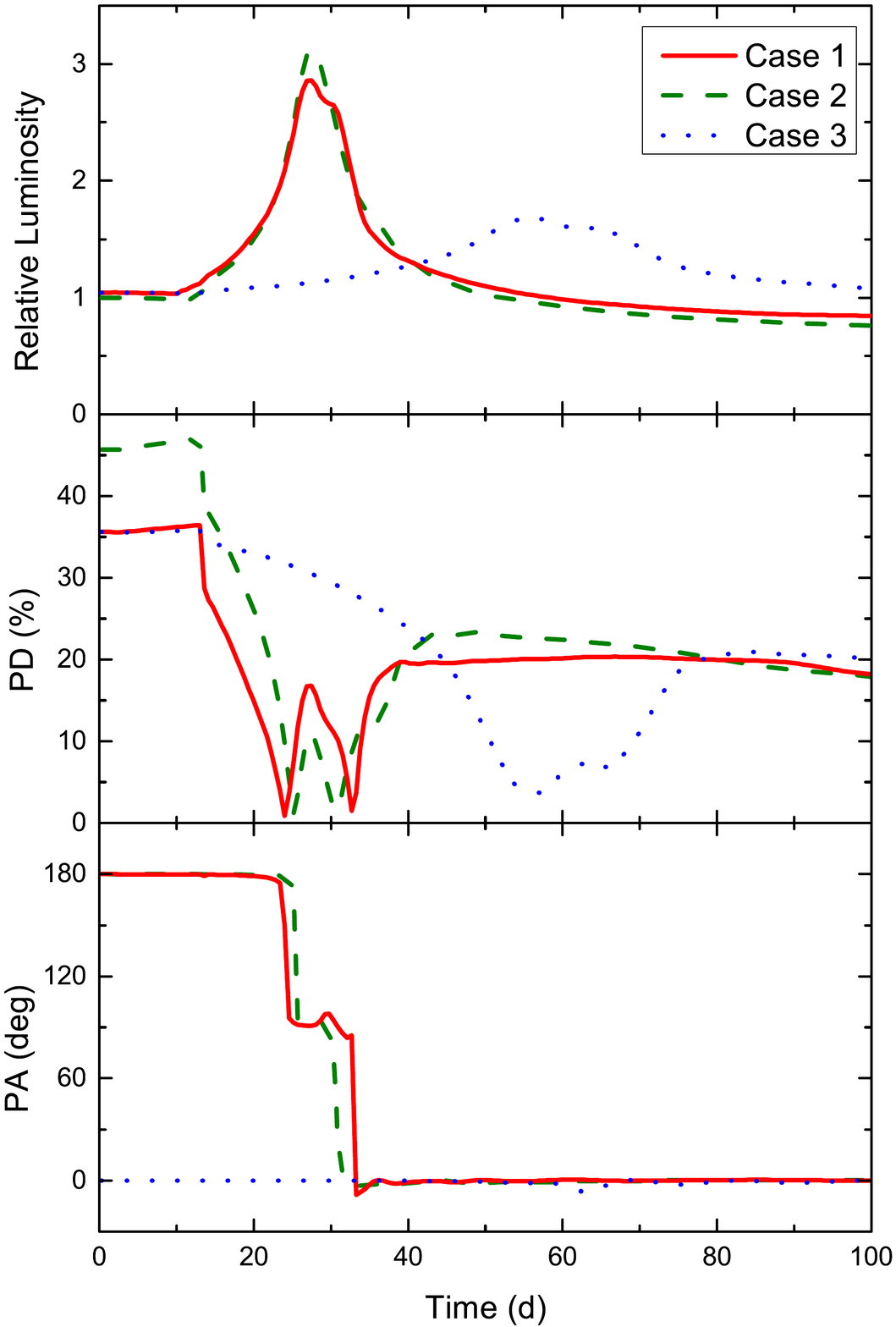}
\caption{Time-dependent light curves as well as PD and PA variations for all cases in days. Upper panel is the relative luminosity, where $1$ is about $10^{45}~{\rm erg\,s^{-1}}$, as in a bright blazar. Middle panel is the PD in percent. Lower panel is the PA in the unit of degrees. Red solid lines are for Case 1, green dashed lines are for Case 2, and blue dotted lines are for Case 3. \label{kink2}}
\end{figure}

Case 3 has the same initial magnetic topology as that in Case 1 but a much lower magnetization, $\sigma_m=0.2$. Hence we observe that the two cases have identical initial quiescent states (Fig. \ref{kink2}). The general kink evolution for Case 3 is similar to that for Case 1, except that the evolution is much slower. Specifically, the linear growth phase in Case 1 takes about $80$ code time units, while that in Case 3 takes $200$ time units. In addition, in Case 1 Phase B starts at $T\sim 50$, but in Case 3 it starts later at $T\sim 120$. Both effects significantly stretch the flare duration and delay the flare peak in Case 3 (Fig. \ref{kink2}). Owing to the much lower magnetic energy conversion rate, the flare level in Case 3 is only $\sim 50\%$, consistent with the injection level. But the slower kink evolution results in that the positive $\vec{j}\cdot \vec{E}$ regions maintain close to the central region for a longer time. As a result, we still observe a significant drop in the PD. Nevertheless, the flare level is not adequately strong to make a PA swing, but only small PA fluctuations at the flare peak. At Phase C, we observe that the smooth helices are weakly distorted (Fig. \ref{case3} lower panel second column), implying that slower kink modes start to surface. This is similar to the evolution in Case 1. However, the injection level is weak compared to $n_{bkg}$, thus the future radiation signatures can only make minor fluctuations, which are of little interest. Notice that the final PD is consistent with that in Case 1.

The major difference of Case 3 from Case 1 is the slower evolution and the lower magnetic energy conversion rate. In our simulation, we do not consider an external energy supply. However, in reality, blazars, for example, can continuously fuel the emission region. In a weakly magnetized emission region, the energy build-up rate can be higher than the conversion rate, thus the magnetization of the emission region may keep increasing. This contributes to a higher synchrotron efficiency as well. In addition, the extra energy can give rise to active variability. Later, the magnetization becomes high enough so that the kink can quickly develop and release the accumulated magnetic energy in the form of flares. This entire process makes an active phase of the emission region, which is frequently observed in blazars.

\section{Implications for Observations \label{implication}}

In this section, we study the general features of the polarized emission intrinsically arising from kink instabilities, and discuss their connection with observations.

We can see in Fig. \ref{kink2} that the light curves and polarization signatures are generally symmetric in time, which are consistent with most observations \citep{Abdo10,Blinov15}. The reason lies in the LTTEs. For the overall shapes of the light curves and polarization signatures, since the kink evolution time scale is comparable to the light crossing time scale, the LTTEs wipe out most asymmetric features, such as the asymmetric shape of the injection level (Fig. \ref{kink1}). This is because the kink evolution time scale is governed by the Alf\'ven speed. In a moderately magnetized environment, the Alf\'ven speed is relativistic. We notice that in Case 3, however, the Alf\'ven speed is non-relativistic, due to the low magnetization. Therefore, the intrinsically time-asymmetric injection profile dominates the radiation signatures. We can observe that the light curve for Case 3 clearly shows a fast rise with a slow decay. Another factor is that the positive $\vec{j}\cdot \vec{E}$ generally distributes symmetric along the central axis of the emission region. In addition, these regions are relatively close to the central axis. Therefore, the polarized emission around the flare peak is also symmetric in time. We notice that in Case 2 after the flare peak the strong injection regions move away from the central axis. In this situation, the polarized emission coming from regions that are far from the observer are delayed due to the LTTEs. This causes a slower recovery of PD after the flare peak for Case 2 (Fig. \ref{kink2}).

By comparing the first two cases and Case 3, it is clear that for kink-initiated flares, stronger flares are likely to accompany with PA swings. This is consistent with many observations \citep[e.g.,][]{Lister05,Marscher08,Abdo10,Blinov15}. The reason is that although the planar magnetic contribution dominates the quiescent state, at the injection peak, most regions with strong injection are around the central axis of the emission region, where the vertical magnetic component is strong. For strong flares, this strong injection with vertical magnetic contribution results in a switch of the polarization dominance, leading to a $90^{\circ}$ PA rotation. At later stage the strong injection regions move away from the central axis, into regions where the planar magnetic component is strong. Hence the PA rotates another $90^{\circ}$ to its initial state. Nevertheless, for weak flares, the injection at the flare peak is not strong enough to dominate over the initial planar domination, hence no PA swing is observed.

Interestingly, although Case 2 starts with a stronger toroidal component, in the end the PD stays at the same value as in Case 1 and 3. This suggests that the kink instability can sufficiently alter the magnetic topology to make comparable vertical and the planar contributions and induce some turbulence, although it still keeps a generally laminar magnetic structure. We notice that the final PD at $\lesssim 20\%$ for all cases naturally explains the general blazar PD during the quiescent state, indicating that kink instabilities may widely exist in the blazar emission region.

\section{Summary and Discussions \label{discussion}}

We have presented the polarization-dependent radiation modeling of the kink instability in the blazar emission region based on RMHD simulations. Our study involves self-consistent magnetic field evolution as well as comprehensive radiation transfer, and nonthermal particle injection based on the local magnetic energy conversion rate derived from RMHD simulations. Therefore, our results represent the intrinsic radiation and polarization signatures originating from the kink instability. We summarize our major findings as follows.

Regarding features in the kink evolution:
\begin{enumerate}
\item In a sufficiently magnetized environment, kink instabilities can efficiently convert the magnetic energy.
\item Multiple kink modes can coexist in the system, but the one with the fastest growth rate generally dominates the magnetic energy conversion.
\item Generally speaking, the magnetic energy conversion first concentrates around the central part of the jet, then moves outwards.
\item The magnetic topology can be sufficiently altered by the kink instability, but it can still remain laminar to some extent.
\end{enumerate}

Regarding observational phenomena, we find that:
\begin{enumerate}
\item The time-dependent radiation and polarization signatures from kink instabilities are generally symmetric in time; however, a low magnetization can cause a slower decay in the light curve, and a strong initial toroidal magnetic component can result in a slower recovery of PD.
\item Stronger flares arise from faster kink evolution, which often leads to PA swings; weaker flares are accompanied by PA fluctuations.
\item The PD at the end of the kink evolution is around $10-20\%$, which is consistent with the PD at quiescent state, implying that kink may widely exist in the blazar emission environment.
\end{enumerate}

\cite{ZHC16a} has performed detailed polarization-dependent radiation modeling of the shock based on RMHD simulations. In that paper, shocks can make similar radiation and polarization signatures to the kink instability. Also the trend that stronger flares are probably accompanied by PA swings is similar. However, there are several major differences. The first one is that in the shock model the nonthermal particle energy comes from the plasma kinetic energy. In order to make strong flares, the magnetization in the emission region is relatively weak. As is shown in \cite{ZHC16a}, shocks can significantly alter the magnetic topology permanently, giving rise to very high PD that is never observed. The second issue is that even if the emission region is considerably magnetized, a strong shock that can double the flux level will also introduce strong polarization variations, so that the final PD may stay at a high level ($\sim 30\%$). This is inconsistent with the commonly observed blazar PD, which is around $10-20\%$. A possible way to resolve these issues is adding a considerable turbulent component. In our current study, however, the general radiation and polarization signatures are mainly within the observational values, even in the case of strong flares. Therefore, we prefer the magnetic-driven flares over the kinetic-driven flares, unless the emission region is strongly turbulent.

In radio observations, people frequently observe bending jets, which are likely resulting from kink instabilities \citep[e.g.,][]{Pearson88,Taylor94,Polatidis95}. Additionally, several papers have suggested that a bending jet or a moving emission blob along a helical path can explain the simultaneous multiwavelength flares and PA swings in blazars \citep{Marscher08,Abdo10,Larionov13}. This requires a helical shape jet structure that can be relatively stable over a long time. Our simulations show that during the nonlinear phase of the kink evolution, the magnetic field lines have been curved into approximately helices, and can maintain over a long period. Therefore, the kink instability naturally provides theoretical supports for the above phenomena/models.

We also notice that the kinked structure appears periodic along the jet direction. If a shock propagates through the kinked jet, it can light up each kink node along its path. Since the kinked jet has released a significant amount of its magnetic energy, the shock may propagate through a number of kink nodes before it dissipates. This may provide a natural explanation of the quasi-periodic oscillation (QPO) signatures that have been reported in a number of sources, such as jetted tidal disruption events (TDEs) and blazars \citep{Valtonen06,Zauderer11}. In particular, many blazar QPO signatures do not repeat for many periods, which are consistent with the dissipation of a shock through a kinked jet. We will study this model in detail in a future paper.

Finally, the radiation and polarization signatures reported here presumes a blazar-like environment, which requires some background nonthermal particle population in the quiescent state. However, in other objects, such as GRBs, a quiescent state emission is not necessary. In such situations, the kink should first make vertical component dominating emission at the beginning of the linear growth phase, then make planar component dominating emission when the kink saturates. This generates a natural $90^{\circ}$ PA swing, and comparable PD in the two phases. This has been reported by \cite{Yonetoku11} in gamma-ray polarimetry of GRBs.

\acknowledgments{HZ, HL, and FG acknowledge support from the LANL/LDRD program and by DoE/Office of Fusion Energy Science through CMSO. HZ and GBT acknowledge support from LANL and from the NASA Fermi Guest Investigator program, grants NNX12A075G, NNX14AQ87G and NNX15AU85G. Simulations are carried out on LANL clusters provided by LANL Institutional Computing.}

\clearpage

\end{document}